\documentclass[preprint,prd]{revtex4}
\usepackage{amsfonts}
\usepackage{amssymb}
\usepackage{amsmath}
\usepackage{graphicx}

\input{tcilatex}
\begin{document}

\title[]{Classical Interaction of a Magnet and a Point Charge: The
Shockley-James Paradox}
\author{Timothy H. Boyer}
\affiliation{Department of Physics, City College of the City University of New York, New
York, New York 10031}

\begin{abstract}
It is pointed out that Coleman and Van Vleck make a major blunder in their
discussion of the Shockly-James paradox by designating relativistic
\textquotedblleft hidden \textit{mechanical} momentum\textquotedblright\ as
the basis for resolution of the paradox. \ This blunder has had a wide
influence in the current physics literature, including erroneous work on the
Shockley-James paradox, on Mansuripur's paradox, on the motion of a magnetic
moment, on the Aharonov-Bohm phase shift, and on the Aharonov-Casher phase
shift. \ Although hidden \textit{mechanical} momentum is indeed dominant for 
\textit{non-interacting} particles moving in a closed orbit under the
influence of an external electric field, the attention directed toward
hidden \textit{mechanical} momentum represents a fundamental
misunderstanding of the classical electromagnetic interaction between a
multiparticle magnet and an external point charge. \ In the \textit{%
interacting} multiparticle situation, the external charge induces an
electrostatic polarization of the magnet which leads to an internal \textit{%
electromagnetic} momentum in the magnet where both the electric and magnetic
fields for the momentum are contributed by the magnet particles. \ This
internal electromagnetic momentum for the interacting multiparticle
situation is equal in magnitude and opposite in direction compared to the
familiar external electromagnetic momentum where the electric field is
contributed by the external charged particle and the magnetic field is that
due to the magnet. \ In the present article, the momentum balance of the
Shockley-James situation for a system of a magnet and a point charge is
calculated in detail for a magnet model consisting of two interacting point
charges which are constrained to move in a circular orbit on a frictionless
ring with a compensating negative charge at the center. \ 
\end{abstract}

\maketitle

\section{Introduction}

The interaction of a magnet and an external point charge has attracted
attention for over a century. \ In 1967, Shockley and James\cite{S-J}
pointed out a paradox for this system; there is an apparent momentum
imbalance between the charge and the magnet when the magnetic moment of the
magnet changes due to some internal forces. \ Thus it was suggested that one
can imagine the magnet to consist of two counter-rotating disks of opposite
charge; the disks are brought into frictional contact and come to rest so
that the magnetic moment is reduced to zero. \ The Faraday induction field
arising from the changing magnetic moment gives an obvious electric force to
change the linear momentum of the external charge; however, the balancing
change in the linear momentum of the magnet is not so obvious. This
\textquotedblleft Shockley-James paradox\textquotedblright\ attracted the
attention of Coleman and Van Vleck\cite{C-VV} who provided a sophisticated
discussion of the situation and directed attention to the relativistic 
\textit{mechanical} momentum of the charge carriers for the magnetic moment.
\ The relativistic "hidden mechanical momentum" emphasized by Coleman and
Van Vleck has been widely adopted. \ In his \textquotedblleft Resource
Letter EM-1:Electromagnetic Momentum,\textquotedblright\ Griffiths lists
twenty-three articles on \textquotedblleft hidden
momentum.\textquotedblright \cite{G-RL} \ \ \textquotedblleft Hidden
mechanical momentum\textquotedblright\ now appears in the electromagnetism
textbooks.\cite{Jackson3rd}\cite{Griffiths3rd} \ It has been used as the
basis for erroneous descriptions of the equation of motion of a magnetic
moment in an electric field,\cite{V} and of the classical electromagnetic
aspects of the Aharonov-Casher phase shift\cite{APV} and the Aharonov-Bohm
phase shift,\cite{ABhm}. \ Most recently the ideas of \textquotedblleft
hidden mechanical momentum\textquotedblright\ were used in connection with
rebuttals\cite{rebut} to Mansuripur's claim\cite{Mansuripur} of the
inconsistency of the Lorentz force law with special relativity. \ 

In the present article, we point out that Coleman and Van Vleck's
"hidden-mechanical-momentum"-solution to the Shockley-James paradox
represents a major blunder. The blunder consists in the claim that the
\textquotedblleft mechanical momentum\textquotedblright\ of the current
carriers of the multiparticle magnet provides the internal momentum in the
magnet (required by Lorentz invariance) which balances the familiar \
electromagnetic momentum arising from the electric field of the external
charge and magnetic field of the magnet. \ 

Indeed, the claim of importance for \textquotedblleft mechanical
momentum\textquotedblright\ in a multiparticle electromagnetic system should
immediately be regarded with suspicion. For a multiparticle electromagnetic
system in a situation where particle collisions are not important, it seems
highly unlikely that the mass of the current carriers is of significance
compared to the electromagnetic interactions between the charges. \ Thus,
for example, the textbook discussions of the self-inductance of a circuit
make no reference whatsoever to the mass of the charges carrying the
currents of the circuit. \ Rather it is the interaction of the accelerating
charges through their electromagnetic fields which overwhelms any mechanical
inertia of the charges' mass, and so determines the acceleration of the
charges.\cite{Ba} \ 

In the present article, we point out that in the interacting-multiparticle
electromagnetic situation, the external charge induces a crucial
electrostatic polarization of the magnet which leads to an internal \textit{%
electromagnetic} momentum in the magnet where both the electric and magnetic
fields for the momentum are contributed by the magnet particles. \ This
internal electromagnetic momentum for the interacting multiparticle
situation is equal in magnitude and opposite in direction compared to the
familiar external electromagnetic field momentum where the electric field is
contributed by the external charged particle and the magnetic field is that
due to the magnet. \ As a transparent illustration of these ideas, we
discuss a situation corresponding in essence to the Shockley-James paradox.
In the present article, for reasons of clarity, \textit{external} forces
(rather than internal frictional forces) are applied which give the magnet a
changing magnetic moment, and therefore an electromagnetic impulse is
delivered to the external charge. \ Momentum conservation is exhibited in
detail for a magnet model consisting of two moving and interacting point
charges of low velocity which are constrained to move in a circular orbit on
a frictionless ring with a balancing negative charge at the center. \ In
this easily-calculated example, the magnet contains internal \textit{%
electromagnetic} momentum, the \textit{total electromagnetic momentum} of
the system vanishes, and the\textit{\ canonical} momentum of the external
charge remains constant (as was noted by Coleman and Van Vleck). \ The
correct understanding of the magnet-point charge interaction has major
implications for our understanding of the motion of a magnetic moment, and
of the experimentally observed Aharonov-Bohm and Aharonov-Casher phase
shifts.

\section{Use of the Darwin Lagrangian Approximation}

\subsection{The Darwin Lagrangian}

Except for the blunder in connection with the final step of their analysis,
Coleman and Van Vleck provide an extremely helpful discussion of the
Shockley-James paradox. \ Here we will follow major portions of their
discussion. \ Coleman and Van Vleck note that the Shockley-James paradox
must be treated through order $1/c^{2}$ (in gaussian units) in the
electromagnetic interactions since the obvious Faraday induction field
acting on the external charge is of order $1/c^{2}.$ \ To this $1/c^{2}$%
-order$,$ the forces between a magnet and a point charge can be understood
in terms of the electromagnetic interactions described by the Darwin
Lagrangian. \ This truncated description of electromagnetism includes all
electromagnetic forces through order $1/c^{2}$ but omits radiation fields
entirely. \ The Darwin Lagrangian for point charges $e_{i}$ with masses $%
m_{i},$ displacements $\mathbf{r}_{i},$ and velocities $\mathbf{v}_{i}$
takes the form\cite{DL}

\begin{eqnarray}
\mathcal{L} &\mathcal{=}&\tsum\limits_{i=1}^{i=N}m_{i}c^{2}\left( -1+\frac{%
\mathbf{v}_{i}^{2}}{2c^{2}}+\frac{(\mathbf{v}_{i}^{2})^{2}}{8c^{4}}\right) -%
\frac{1}{2}\tsum\limits_{i=1}^{i=N}\tsum\limits_{j\neq i}\frac{e_{i}e_{j}}{|%
\mathbf{r}_{i}-\mathbf{r}_{j}|}  \notag \\
&&+\frac{1}{2}\tsum\limits_{i=1}^{i=N}\tsum\limits_{j\neq i}\frac{e_{i}e_{j}%
}{2c^{2}}\left[ \frac{\mathbf{v}_{i}\cdot \mathbf{v}_{j}}{|\mathbf{r}_{i}-%
\mathbf{r}_{j}|}+\frac{\mathbf{v}_{i}\cdot (\mathbf{r}_{i}-\mathbf{r}_{j})%
\mathbf{v}_{j}\cdot (\mathbf{r}_{i}-\mathbf{r}_{j})}{|\mathbf{r}_{i}-\mathbf{%
r}_{j}|^{3}}\right]  \label{Darwin}
\end{eqnarray}

\subsection{Equations of Motion}

The Lagrangian equations of motion derived from Eq. (\ref{Darwin}) can be
rewritten in Newtonian form with the force given by the Lorentz force due to
the other charges%
\begin{eqnarray}
&&\frac{d\mathbf{p}_{i}^{mechanical}}{dt}=\frac{d}{dt}\left[ m_{i}\gamma _{i}%
\mathbf{v}_{i}\right] =\frac{d}{dt}\left[ \frac{m_{i}\mathbf{v}_{i}}{(1-%
\mathbf{v}_{i}^{2}/c^{2})^{1/2}}\right] \approx \frac{d}{dt}\left[
m_{i}\left( 1+\frac{\mathbf{v}_{i}^{2}}{2c^{2}}\right) \mathbf{v}_{i}\right]
\notag \\
&=&e_{i}\tsum\limits_{j\neq i}\mathbf{E}_{j}(\mathbf{r}_{i},t)+e_{i}\frac{%
\mathbf{v}_{i}}{c}\times \tsum\limits_{j\neq i}\mathbf{B}_{j}(\mathbf{r}%
_{i},t)
\end{eqnarray}%
The electric field due to the $j$th particle is given through order $1/c^{2}$
by\cite{PA}%
\begin{eqnarray}
\mathbf{E}_{j}(\mathbf{r,t}) &=&e_{j}\frac{(\mathbf{r}-\mathbf{r}_{j})}{|%
\mathbf{r}-\mathbf{r}_{j}|^{3}}\left[ 1+\frac{\mathbf{v}_{j}^{2}}{2c^{2}}-%
\frac{3}{2}\left( \frac{\mathbf{v}_{j}\cdot (\mathbf{r}-\mathbf{r}_{j})}{c|%
\mathbf{r}-\mathbf{r}_{j}|}\right) ^{2}\right]  \notag \\
&&-\frac{e_{j}}{2c^{2}}\left( \frac{\mathbf{a}_{j}}{|\mathbf{r}-\mathbf{r}%
_{j}|}+\frac{\mathbf{a}_{j}\cdot (\mathbf{r}-\mathbf{r}_{j})(\mathbf{r}-%
\mathbf{r}_{j})}{|\mathbf{r}-\mathbf{r}_{j}|^{3}}\right)  \label{EF}
\end{eqnarray}%
and the magnetic field is 
\begin{equation}
\mathbf{B}_{j}(\mathbf{r},t)=e_{j}\frac{\mathbf{v}_{j}}{c}\times \frac{(%
\mathbf{r}-\mathbf{r}_{j})}{|\mathbf{r}-\mathbf{r}_{j}|^{3}}
\end{equation}%
where the quantity $\mathbf{a}_{j}$ in Eq. (\ref{EF}) corresponds to the
acceleration of the $j$th particle.

\subsection{Canonical Linear Momentum from the Darwin Lagrangian}

Following from the Darwin Lagrangian in Eq. (\ref{Darwin}), we find the
canonical linear momentum of the $i$th charge 
\begin{equation}
\frac{\partial \mathcal{L}}{\partial \mathbf{v}_{i}}=\mathbf{p}%
_{i}^{canonical}=m_{i}\left( 1+\frac{\mathbf{v}_{i}^{2}}{2c^{2}}\right) 
\mathbf{v}_{i}+\tsum\limits_{j\neq i}\frac{e_{i}e_{j}}{2c^{2}}\left( \frac{%
\mathbf{v}_{j}}{|\mathbf{r}_{i}-\mathbf{r}_{j}|}+\frac{\mathbf{v}_{j}\cdot (%
\mathbf{r}_{i}-\mathbf{r}_{j})(\mathbf{r}_{i}-\mathbf{r}_{j})}{|\mathbf{r}%
_{i}-\mathbf{r}_{j}|^{3}}\right)
\end{equation}%
The canonical linear momentum includes a \textit{mechanical} linear momentum
of the $i$th particle%
\begin{equation}
\mathbf{p}_{i}^{mechanical}=m_{i}\gamma _{i}\mathbf{v}_{i}\approx
m_{i}[1+v_{i}^{2}/(2c^{2})]\mathbf{v}_{i}=m_{i}\mathbf{v}_{i}+m_{i}v_{i}^{2}%
\mathbf{v}_{i}/(2c^{2})
\end{equation}%
and also the \textit{electromagnetic} linear momenta $\mathbf{p}_{i}^{em}$\
associated with the \textit{electric} field $\mathbf{E}_{i}$ of the $i$th
particle and \textit{magnetic} fields $\mathbf{B}_{j},$ $j\neq i,$\ of all
the other particles%
\begin{equation}
\mathbf{p}_{i}^{em}=\dsum\limits_{j\neq i}\frac{1}{4\pi c}\tint d^{3}r%
\mathbf{E}_{i}\times \mathbf{B}_{j}=\tsum\limits_{j\neq i}\frac{e_{i}e_{j}}{%
2c^{2}}\left( \frac{\mathbf{v}_{j}}{|\mathbf{r}_{i}-\mathbf{r}_{j}|}+\frac{%
\mathbf{v}_{j}\cdot (\mathbf{r}_{i}-\mathbf{r}_{j})(\mathbf{r}_{i}-\mathbf{r}%
_{j})}{|\mathbf{r}_{i}-\mathbf{r}_{j}|^{3}}\right)
\end{equation}

\section{Vanishing Total Momentum for a Point Charge at Rest Outside a Magnet%
}

\subsection{Vanishing Total Momentum from the Requirement of Lorentz
Invariance}

Coleman and Van Vleck provide a sophisticated proof (based on Lorentz
invariance) that the total momentum of a magnet-point charge system must
vanish when held at steady state. \ In the present article, we are
interested in providing physical insight into the aspect that Coleman and
Van Vleck missed. \ Thus, rather than providing a general discussion, we
will introduce a simple model for the magnet which we will use later in
explicit calculations.

We consider the situation of a point charge $q$ at rest outside a
multiparticle magnet which is modeled as $N$ point charges $e_{i},$ $%
i=1,2,...N,$ of equal charge $e$ and mass $m$, moving without friction with
a steady current in a circle of radius $R$ about the origin in the $xy-$%
plane. \ There is a negative point particle of charge $-Ne$ at the center of
this circle. \ Thus the magnetic may be pictured as a set of beads of equal
positive charge $e$ sliding on a frictionless ring with a balancing negative
charge at the center of the ring. \ The external point charge $q$ will cause
an electrostatic polarization of the magnet so that the spacing between the
charges $e_{i}$ will vary around the circle. \ In the
interacting-multiparticle case, the masses of the charges $e$ will be of no
significance, and the charges will move through the electrostatic charge
distribution produced by the external charge $q.$ \ There is an external
force of constraint applied to the charge $q$ which keeps $q$ at rest
despite the electrostatic attraction. \ This electrostatic force must be
along the line $\mathbf{r}_{q}$ running from the center of the magnet to the
charge $q.~$\ There are also radial forces of constraint applied to the
magnet charges so that they remain in the circular orbit of radius $R$. \
The center-of-energy conservation law for a relativistic theory then requires%
\cite{CE}%
\begin{equation}
\tsum\limits_{i=1}^{N}(\mathbf{F}_{\text{ext}i}\cdot \mathbf{v}_{i})\mathbf{r%
}_{i}=\frac{d}{dt}(U\overrightarrow{X}_{CE})-c^{2}\mathbf{P}_{system}
\label{CE}
\end{equation}%
so that the power delivered by the external forces $\mathbf{F}_{\text{ext}i}$
weighted by the location $\mathbf{r}_{i}$ of the power delivery equals the
time-rate-of-change of the system energy $U$ times the center-of-energy $%
\overrightarrow{X}_{CE}$ minus $c^{2}$ times the total system momentum $%
\mathbf{P}_{system}$. \ For a real magnet, there are no localized
distributed sources of power, and indeed, in our model, there is no power
being introduced by the radial forces of constraint; the moving particles of
the magnet move perpendicular to the forces of constraint for the circular
orbit. \ In steady state, the energy times the center of energy is not
changing in time, and therefore it follows from Eq. (\ref{CE}) that 
\begin{equation}
\mathbf{P}_{system}=0  \label{P0}
\end{equation}%
\ the total system momentum (of the magnet and the external charge) must
vanish.

\subsection{Familiar External Electromagnetic Momentum}

Now it is a familiar textbook observation that a steady localized current in
the presence of an external electric field leads to linear momentum in the
electromagnetic field.\cite{Jackson6.5} \ In the present case, this
electromagnetic field momentum $\mathbf{P}_{q-\mu }^{em}$ involves the
electric field $\mathbf{E}_{q}$ of the external charge $q$ and the magnetic
field $\mathbf{B}_{\mu }$ of the moving charges of the magnet%
\begin{equation}
\mathbf{P}_{q-\mu }^{em}=\frac{1}{4\pi c}\tint d^{3}r\,\mathbf{E}_{q}\times 
\mathbf{B}_{\mu }\mathbf{=}\tsum\limits_{i=1}^{N}\frac{qe}{2c^{2}}\left( 
\frac{\mathbf{v}_{i}}{|\mathbf{r}_{i}-\mathbf{r}_{q}|}+\frac{\mathbf{v}%
_{i}\cdot (\mathbf{r}_{i}-\mathbf{r}_{q})(\mathbf{r}_{i}-\mathbf{r}_{q})}{|%
\mathbf{r}_{i}-\mathbf{r}_{q}|^{3}}\right)   \label{Pqmu}
\end{equation}%
If the charge $q$ at displacement $\mathbf{r}_{q}$ is far from the magnet of
radius $R$, $R<<r_{q},$ then we can expand this expression in powers of $%
R/r_{q}.$ This momentum in Eq. (\ref{Pqmu}) then becomes%
\begin{eqnarray}
\mathbf{P}_{q-\mu }^{em} &\approx &\tsum\limits_{i=1}^{N}\frac{qe}{2c^{2}}%
\left[ \frac{\mathbf{v}_{i}}{r_{q}}\left( 1+\frac{\mathbf{r}_{i}\cdot 
\mathbf{r}_{q}}{r_{q}^{2}}\right) +\frac{(\mathbf{v}_{i}\cdot \mathbf{r}_{i}-%
\mathbf{v}_{i}\cdot \mathbf{r}_{q})(\mathbf{r}_{i}-\mathbf{r}_{q})}{r_{q}^{3}%
}\left( 1+3\frac{\mathbf{r}_{i}\cdot \mathbf{r}_{q}}{r_{q}^{2}}\right) %
\right]   \notag \\
&=&\frac{qe}{2c^{2}}\tsum\limits_{i=1}^{N}\left\{ \left( \frac{\mathbf{v}_{i}%
}{r_{q}}+\frac{(\mathbf{v}_{i}\cdot \mathbf{r}_{q})\mathbf{r}_{q}}{r_{q}^{3}}%
\right) +\left( \frac{-(\mathbf{v}_{i}\cdot \mathbf{r}_{i})\mathbf{r}_{q}}{%
r_{q}^{3}}+\frac{3(\mathbf{v}_{i}\cdot \mathbf{r}_{q})(\mathbf{r}_{i}\cdot 
\mathbf{r}_{q})\mathbf{r}_{q}}{r_{q}^{5}}\right) \right.   \notag \\
&&\left. +\left( \frac{(\mathbf{r}_{i}\cdot \mathbf{r}_{q})\mathbf{v}_{i}-(%
\mathbf{v}_{i}\cdot \mathbf{r}_{q})\mathbf{r}_{i}}{r_{q}^{3}}\right)
\right\} +O(1/r_{q}^{3})  \notag \\
&=&\frac{1}{c}\left( \frac{e}{2c}\tsum\limits_{i=1}^{N}\mathbf{r}_{i}\times 
\mathbf{v}_{i}\right) \times \frac{q\mathbf{r}_{q}}{r_{q}^{3}}=\frac{1}{c}%
\overrightarrow{\mu }\times \frac{q\mathbf{r}_{q}}{r_{q}^{3}}=\frac{-1}{c}%
\overrightarrow{\mu }\times \mathbf{E}_{q}(0)  \label{Pemqmu}
\end{eqnarray}%
where $\overrightarrow{\mu }$ is the magnetic moment of the magnet, $\mathbf{%
E}_{q}(0)=-q\mathbf{r}_{q}/r_{q}^{2},$ and where we have used $\tsum \mathbf{%
v}_{i}=(d/dt)\tsum \mathbf{r}_{i}=0,$ $\tsum \mathbf{v}_{i}\cdot \mathbf{r}%
_{i}=(d/dt)\tsum r_{i}^{2}/2=0,$ and $\tsum (\mathbf{v}_{i}\cdot \mathbf{r}%
_{q})(\mathbf{r}_{i}\cdot \mathbf{r}_{q})=(d/dt)\tsum (\mathbf{r}_{i}\cdot 
\mathbf{r}_{q})(\mathbf{r}_{i}\cdot \mathbf{r}_{q})/2=0$ for a steady-state
magnet current. This result gives the electromagnetic field momentum $%
\mathbf{P}_{q-\mu }^{em}$ which clearly must be balanced by some other
momentum so as to meet the relativistic restriction we found above in Eq. (%
\ref{P0}) that the total system momentum vanishes, $\mathbf{P}_{system}=0.$
\ Coleman and Van Vleck were aware of the single-charge current-loop model
of a magnet,\cite{MIT}, and, without carrying out any detailed calculation,
they identified the missing piece of the total momentum as mechanical
momentum associated with the motion of the masses of the current carriers in
the magnet.\cite{C-VVftnt} \ Such an identification is a blunder for an
interacting-multiparticle magnet. \ 

\subsection{Vanishing Total \textit{Electromagnetic} Momentum for an \textit{%
Interacting}-\textit{Multiparticle} Magnet}

Let us go back and reconsider the standard textbook calculation of the
electromagnetic field momentum for a magnet-point charge system.\cite%
{Jackson6.5} \ For a localized steady-state situation, the total
electromagnetic field momentum is given by 
\begin{equation}
\mathbf{P}_{total}^{em}=\frac{1}{4\pi c}\tint d^{3}r\,\mathbf{E}\times 
\mathbf{B}=\frac{-1}{4\pi c}\tint d^{3}r\,\mathbf{\nabla }\Phi \times 
\mathbf{B=}\frac{1}{4\pi c}\tint d^{3}r\,\Phi \mathbf{\nabla }\times \mathbf{%
B}=\frac{1}{c^{2}}\tint d^{3}r\Phi \mathbf{J}  \label{Ptot}
\end{equation}%
where $\mathbf{E=-\nabla }\Phi ,$ and we have used the assumption that both
the charge and current distributions are localized so as to drop the surface
term involved in the partial integration. In the textbook example, it is
assumed that the only contribution to the electrostatic potential $\Phi $
comes from the external charge $q,$%
\begin{equation}
\Phi (\mathbf{r})=\Phi _{q}(\mathbf{r})=\frac{q}{|\mathbf{r}-\mathbf{r}_{q}|}
\label{PHI}
\end{equation}%
However, for a multiparticle magnet, the speed of the charges will become
ever smaller for a fixed magnetic moment as the number of charges is
increased. \ Then we expect that the external charge $q$ will cause a
polarization of the magnet which is approximately that of electrostatics. \
Now if the charge $q$ produces an electrostatic polarization of the magnet,
then there is a second source of the electrostatic potential, that due to
the electrostatic charge distribution of the magnet. \ Therefore instead of
Eq. (\ref{PHI}), we must write%
\begin{equation}
\Phi (\mathbf{r})=\frac{q}{|\mathbf{r}-\mathbf{r}_{q}|}+\left(
\tsum\limits_{i=1}^{N}\frac{e}{|\mathbf{r}-\mathbf{r}_{i}|}+\frac{-Ne}{|%
\mathbf{r}|}\right)
\end{equation}%
The first term on the right-hand side is the electrostatic potential $\Phi
_{q}(\mathbf{r})$\ of the external charge $q$, and the second term in the
round brackets includes the electrostatic potential of the moving magnet
charges $e_{i}$ $\ $together with the potential of the central negative
charge. \ Now the only current density $\mathbf{J}$ is that associated with
the moving charges of the magnet 
\begin{equation}
\mathbf{J}(\mathbf{r},t)=\tsum\limits_{i=1}^{N}e\mathbf{v}_{i}\delta ^{3}[%
\mathbf{r}-\mathbf{r}_{i}(t)]
\end{equation}%
Thus the total electromagnetic field momentum in Eq. (\ref{Ptot}) is given
by 
\begin{equation}
\mathbf{P}_{total}^{em}=\frac{1}{c^{2}}\tsum\limits_{i=1}^{N}e\mathbf{v}%
_{i}(t)\Phi (\mathbf{r}_{i},t)=\frac{1}{c^{2}}\tsum\limits_{i=1}^{N}e\mathbf{%
v}_{i}(t)\,const=0
\end{equation}%
Here we have used two facts. \ First, in electrostatic equilibrium for the
multiparticle situation, the tangential component of the electric field at
each particle vanishes so that the electrostatic potential along the magnet
is a constant. \ Second, in steady state, the current is constant, so that $%
\tsum e\mathbf{v}_{i}=0.$

Thus we find the situation that the total electromagnetic momentum of the
entire system vanishes $\mathbf{P}_{total}^{em}=0$ because the
electromagnetic field momentum $\mathbf{P}_{q-\mu }^{em}$ (arising from the
electric field of the charge $q$ and magnetic field of the magnet) is
balanced by the electromagnetic momentum $\mathbf{P}_{\mu }^{em}$ (arising
from the induced electrostatic charge distribution of the magnet and the
magnetic fields of the charges of the magnet). \ This second electromagnetic
momentum $\mathbf{P}_{\mathbf{\mu }}^{em}$ involves only charges of the
magnet and can be regarded as momentum internal to the magnet. \ 

In an earlier article, the internal electromagnetic momentum was calculated
and exhibited explicitly for the simple case of a two-moving-particle magnet.%
\cite{Myth} \ This two-particle magnet was treated for moving charges at
finite velocity (retaining the squares of the velocity), and therefore both
internal mechanical momentum and internal electromagnetic momentum are
present in the magnet and contribute to the total momentum required by
relativity in Eq. (\ref{P0}). \ It is only in the multiparticle case or the
low-velocity few-particle case that the internal \textit{electromagnetic}
momentum is dominant and that the internal \textit{mechanical} momentum is
negligible. \ It is only in the multiparticle case or the low-velocity
few-particle case that the induced charge distribution for the magnet goes
over to the electrostatic distribution and the total \textit{electromagnetic}
momentum vanishes. \ 

\subsection{Remarks on Electromagnetic Momentum in the Physics Literature}

The evaluation of the electromagnetic field momentum for a magnet-point
charge system in both the textbook and research literature usually makes the
assumption (often not stated explicitly) that the currents of the magnet are
held at their unperturbed values despite the presence of the external charge.%
\cite{1973} \ Such unperturbed currents require that the external forces
(which hold the currents to their unperturbed values) deliver energy locally
at one part of the current distribution and remove energy locally at a
different part of the current distribution so that there is a net flow of
power and an associated net momentum density.\cite{Furry}\cite{Johnson}\cite%
{Comm} \ Such local, power-delivering external forces do not occur in nature.

It has been remarked that if the magnet is shielded by a closed conductor,
then the total electromagnetic field momentum will vanish since the external
electric field of the charge $q$ would be screened out of the region of the
currents.\cite{Furry}\cite{Johnson} \ However, this has been regarded as a
special situation. \ We suggest that the vanishing of the total
electromagnetic field momentum is the general situation which holds in all
cases involving an interacting-multiparticle magnet which has no local
power-delivering external forces.

\section{Electromagnetic Momentum for a Two-Moving-Interacting-Particle
Magnet}

\subsection{The Two-Moving-Interacting-Particle Model for a Magnet}

In order to emphasize the vanishing total electromagnetic momentum for the
situation of a multiparticle magnet and a point charge, we will carry out
the explicit calculation for a two-moving-interacting-particle magnet
following the model mentioned above. \ Since this is a few-particle magnet,
we must work in the low-velocity limit where we ignore terms in $v_{i}^{2},$
the squares of the velocities of the moving charges in the magnet.

We consider two charges $e_{i},$ $i=1,2,$ of equal charge $e$ located at 
\begin{equation}
\mathbf{r}_{i}=R\left( \widehat{i}\cos \phi _{i}+\widehat{j}\sin \phi
_{i}\right) =R\widehat{r}_{i}
\end{equation}%
in the $xy$-plane, which are free to move on a frictionless circular ring of
radius $R\,$\ about the coordinate origin with a negative charge $-2e$ in
the center. When there is no external charge $q$ present, the steady-state
motion of the charges is given by unperturbed values, 
\begin{equation}
\phi _{0i}=\omega _{0}t+i\pi +\theta _{0},\text{ \ \ \ }i=1,2
\end{equation}%
where $\theta _{0}$ is an initial phase. \ In our calculations, we will
average over the initial phase $\theta _{0}.$ \ The tangential direction $%
\widehat{\phi }_{i}$ for the $i$th particle is given by 
\begin{equation*}
\widehat{\phi }_{i}=-\widehat{i}\sin \phi _{i}+\widehat{j}\cos \phi _{i}
\end{equation*}

\subsection{Perturbation of the Particle Motion Due to the External Charge $%
q $}

The external charge $q$ is located at $\mathbf{r}_{q}$ in the $xy$-plane
outside the magnet, $R<<r_{q},$%
\begin{equation}
\mathbf{r}_{q}=r_{q}\left( \widehat{i}\cos \phi _{q}+\widehat{j}\sin \phi
_{q}\right) =r_{q}\widehat{r}_{q}
\end{equation}%
and 
\begin{equation}
\widehat{\phi }_{q}=\left( -\widehat{i}\sin \phi _{q}+\widehat{j}\cos \phi
_{q}\right) 
\end{equation}%
\ We assume that the two charges are are moving slowly so that they are
nearly in electrostatic equilibrium, with the average angle, $(\phi
_{1}+\phi _{2})/2=\omega _{0}t+const,$ moving at constant angular velocity $%
\omega _{0}.$ \ Due to the presence of the external charge $q$, the charges
are displaced from the unperturbed angular positions $\phi _{0i}$ over to 
\begin{equation}
\phi _{i}=\phi _{0i}+\eta _{i}
\end{equation}%
so that $\mathbf{r}_{i}=\widehat{r}_{i}R=$ $\widehat{i}R\cos (\phi
_{0i}+\eta _{i})+\widehat{j}R\sin (\phi _{0i}+\eta _{i}).$ \ Through first
order in the perturbation caused by the charge $q,$ we find 
\begin{equation}
\widehat{r}_{i}=\widehat{i}\cos (\phi _{0i}+\eta _{i})+\widehat{j}\sin (\phi
_{0i}+\eta _{i})\approx \widehat{i}[\cos \phi _{0i}-\eta _{i}\sin \phi
_{0i}]+\widehat{j}[\sin \phi _{0i}+\eta _{i}\cos \phi _{0i}]=\widehat{r}%
_{0i}+\widehat{\phi }_{0i}\eta _{i}.
\end{equation}%
using the first-order approximations $\sin \eta _{i}\approx \eta _{i}$ and $%
\cos \eta _{i}\approx 1.$ \ We will also need the velocity $\mathbf{v}_{i}$
of the $i$th particle, given by%
\begin{equation}
\mathbf{v}_{i}=\frac{d\mathbf{r}_{i}}{dt}=\widehat{\phi }_{i}v_{i}=R\left(
\omega _{0}+\frac{d\eta _{i}}{dt}\right) [-\widehat{i}\sin (\phi _{0i}+\eta
_{i})+\widehat{j}\cos (\phi _{0i}+\eta _{i})]
\end{equation}%
The unit vector $\widehat{\phi }_{i}$ in the direction of the velocity can
be approximated through first order\ in the perturbation as%
\begin{eqnarray}
\widehat{\phi }_{i} &=&-\widehat{i}\sin (\phi _{0i}+\eta _{i})+\widehat{j}%
\cos (\phi _{0i}+\eta _{i})\approx -\widehat{i}[\sin \phi _{0i}+\eta
_{i}\cos \phi _{0i}]+\widehat{j}[\cos \phi _{0i}-\eta _{i}\sin \phi _{0i}] 
\notag \\
&=&\widehat{\phi }_{0i}-\widehat{r}_{0i}\eta _{i}  \label{phiD}
\end{eqnarray}%
and the particle velocity is approximated as%
\begin{equation}
\mathbf{v}_{i}=R\left( \omega _{0}+\frac{d\eta _{i}}{dt}\right) \widehat{%
\phi }_{0i}-\omega _{0}R\eta _{i}\widehat{r}_{0i}  \label{vel}
\end{equation}%
where $\omega _{0}R=v_{0}$ is the speed of the unperturbed charges.

When the unperturbed angular velocity $\omega _{0}$ is very small, we may
use the nonrelativistic form of Newton's second law for the motion of the
charged particles and\ use the electrostatic fields for the forces. \ Thus
the tangential acceleration of the $i$th particle is given by%
\begin{eqnarray}
mR\frac{d^{2}\phi _{i}}{dt^{2}} &=&mR\frac{d^{2}\eta _{i}}{dt^{2}}=\widehat{%
\phi }_{i}\cdot \lbrack e\mathbf{E}_{q}\mathbf{(r}_{i})+e\mathbf{E}_{j\neq
i}(\mathbf{r}_{i})]=\widehat{\phi }_{i}\cdot \left( \frac{-eq\widehat{r}_{q}%
}{r_{q}^{2}}+\frac{e^{2}(\mathbf{r}_{i}-\mathbf{r}_{j\neq i})}{|\mathbf{r}%
_{i}-\mathbf{r}_{j\neq i}|^{3}}\right)  \notag \\
&=&e\left( -q\frac{\widehat{\phi }_{i}\cdot \widehat{r}_{q}}{r_{q}^{2}}-e%
\frac{\widehat{\phi }_{i}\cdot \widehat{r}_{j\neq i}R}{|\mathbf{r}_{i}-%
\mathbf{r}_{j\neq i}|^{3}}\right) \approx e\left( -q\frac{\widehat{\phi }%
_{0i}\cdot \widehat{r}_{q}}{r_{q}^{2}}-e\frac{(\eta _{i}-\eta _{j\neq i})R}{%
(2R)^{3}}\right)  \label{acc}
\end{eqnarray}%
through first order in the perturbation, where we have used $\widehat{\phi }%
_{i}\cdot \widehat{r}_{i}=0$ and 
\begin{eqnarray}
\widehat{\phi }_{i}\cdot \widehat{r}_{j\neq i} &=&[-\widehat{i}\sin (\phi
_{0i}+\eta _{i})+\widehat{j}\cos (\phi _{0i}+\eta _{i})]\cdot \lbrack 
\widehat{i}\cos (\phi _{0j\neq i}+\eta _{j\neq i})+\widehat{j}\sin (\phi
_{0j\neq i}+\eta _{j\neq i})]  \notag \\
&=&\sin (\phi _{0j\neq i}+\eta _{j\neq i}-\phi _{0i}-\eta _{i})=\sin (\eta
_{i}-\eta _{j\neq i})\approx \eta _{i}-\eta _{j\neq i}  \label{phir}
\end{eqnarray}%
since $\phi _{01}-\phi _{02}=\pi .$ The equation (\ref{acc}) is odd under
the interchange of the two particles. \ Then for the steady state situation,
we must have $\eta _{1}=-\eta _{2},$ and so equation (\ref{acc}) becomes

\begin{equation}
mR\frac{d^{2}\eta _{i}}{dt^{2}}=e\left( -q\frac{\widehat{\phi }_{0i}\cdot 
\widehat{r}_{q}}{r_{q}^{2}}-e\frac{2\eta _{i}R}{(2R)^{3}}\right)
\end{equation}%
Then from the unperturbed motions $\phi _{0i}=\omega _{0}t+i\pi +\theta
_{0}, $ we find the steady-state angular perturbation given by 
\begin{equation}
\eta _{i}=e\frac{-q}{r_{q}^{2}}\widehat{r}_{q}\cdot \widehat{\phi }_{0i}%
\frac{1}{[-mR\omega _{0}^{2}+e^{2}/(2R)^{2}]}  \label{eta}
\end{equation}%
Now here we are interested only in the low-velocity limit, which will
correspond to multiparticle behavior where the electric charge distribution
is essentially the electrostatic distribution. \ Therefore we will retain
terms only through first order in the particle speed $v_{0i}=R\omega _{0}.$
\ Accordingly, \ for $mR^{2}\omega _{0}^{2}<<e^{2}/R,$ the angular
perturbation in Eq. (\ref{eta}) becomes%
\begin{equation}
\eta _{i}=e\frac{-q}{r_{q}^{2}}\widehat{r}_{q}\cdot \widehat{\phi }_{0i}%
\frac{1}{[e^{2}/(2R)^{2}]}=\frac{-4qR^{2}}{er_{q}^{2}}\widehat{r}_{q}\cdot 
\widehat{\phi }_{0i}=\frac{4qR^{2}}{er_{q}^{2}}\sin (\phi _{0i}-\phi _{q})
\label{eta2}
\end{equation}%
which is the same as the electrostatic perturbation. \ Taking the time
derivative in Eq. (\ref{eta2}) and noting $d\widehat{\phi }_{0i}/dt=-\omega
_{0}\widehat{r}_{0i},$ we have%
\begin{equation}
\frac{d\eta _{i}}{dt}=\frac{4q\omega _{0}R^{2}}{er_{q}^{2}}\widehat{r}%
_{q}\cdot \widehat{r}_{0i}=\frac{4q\omega _{0}R^{2}}{er_{q}^{2}}\cos (\phi
_{0i}-\phi _{q})  \label{etaD}
\end{equation}

\subsection{Vanishing Total Electromagnetic Linear Momentum}

When we average over the initial angle $\theta _{0}$ and carry out our
calculation in the low-velocity approximation where we ignore terms in $%
v_{i}^{2},$ then the two-particle magnet behaves as though the circle of
orbital motion is an equipotential and hence the total electromagnetic
momentum of the system vanishes. \ The familiar external electromagnetic
momentum is that given in Eq. (\ref{Pemqmu}). \ The internal electromagnetic
linear momentum is that arising from electric and magnetic fields of the
charges of the magnet alone. \ Thus here the internal electromagnetic
momentum $\mathbf{P}_{\mu }^{em}$\ of the magnet is given (through first
order in the perturbation) by%
\begin{eqnarray}
\mathbf{P}_{\mu }^{em} &=&\tsum\limits_{i=1}^{2}\tint d^{3}r\frac{1}{4\pi c}%
\mathbf{E}_{i}\times \mathbf{B}_{j\neq i}=\tsum\limits_{i=1}^{2}\frac{e^{2}}{%
2c^{2}}\left( \frac{\mathbf{v}_{j\neq i}}{|\mathbf{r}_{i}-\mathbf{r}_{j\neq
i}|}+\frac{\mathbf{v}_{j\neq i}\cdot (\mathbf{r}_{i}-\mathbf{r}_{j\neq i})(%
\mathbf{r}_{i}-\mathbf{r}_{j\neq i})}{|\mathbf{r}_{i}-\mathbf{r}_{j\neq
i}|^{3}}\right)  \notag \\
&\approx &\tsum\limits_{i=1}^{2}\frac{e^{2}}{2c^{2}}\left( \frac{\mathbf{v}%
_{j\neq i}}{2R}+\frac{(v_{j\neq i}\widehat{\phi }_{j\neq i}\cdot \widehat{r}%
_{i})(\widehat{r}_{i}-\widehat{r}_{j\neq i})R^{2}}{(2R)^{3}}\right)
\label{Pemu0}
\end{eqnarray}%
Now using Eq. (\ref{phir}), this becomes through first order in the
perturbation%
\begin{eqnarray}
\mathbf{P}_{\mu }^{em} &\approx &\tsum\limits_{i=1}^{2}\frac{e^{2}}{2c^{2}}%
\left( \frac{\mathbf{v}_{j\neq i}}{2R}+\frac{v_{j\neq i}(\eta _{j\neq
i}-\eta _{i})(\widehat{r}_{i}-\widehat{r}_{j\neq i})R^{2}}{(2R)^{3}}\right) 
\notag \\
&\approx &\tsum\limits_{i=1}^{2}\frac{e^{2}}{2c^{2}}\left( \frac{\mathbf{v}%
_{j\neq i}}{2R}+\frac{v_{0}(-2\eta _{i})(2\widehat{r}_{0i})R^{2}}{(2R)^{3}}%
\right)  \label{Pemu}
\end{eqnarray}%
Now averaging over the initial angle $\theta _{0},$ we note that $%
\left\langle \sin \theta _{0}\right\rangle =\left\langle \cos \theta
_{0}\right\rangle =\left\langle \sin \theta _{0}\cos \theta
_{0}\right\rangle =0,$ while $\left\langle \sin ^{2}\theta _{0}\right\rangle
=\left\langle \cos ^{2}\theta _{0}\right\rangle =1/2.$ $\ $From Eq. (\ref%
{vel}) we have $\left\langle \mathbf{v}_{i}\right\rangle =\left\langle 
\mathbf{v}_{j\neq i}\right\rangle =0$. \ Also, $\left\langle \eta _{i}%
\widehat{r}_{0i}\right\rangle =\widehat{\phi }_{q}2qR^{2}/(er_{q}^{2}).$ \
Therefore Eq. (\ref{Pemu}) becomes 
\begin{eqnarray}
\left\langle \mathbf{P}_{\mu }^{em}\right\rangle &\approx
&\tsum\limits_{i=1}^{2}\frac{e^{2}}{2c^{2}}\left( \frac{\left\langle \mathbf{%
v}_{j\neq i}\right\rangle }{2R}-\frac{v_{0}\left\langle \eta _{i}\widehat{r}%
_{0i}\right\rangle }{2R}\right)  \notag \\
&=&\tsum\limits_{i=1}^{2}\frac{e^{2}v_{0}}{2c^{2}}\left( \frac{4qR^{2}}{%
er_{q}^{2}}\right) \left[ \frac{0}{2R}-\frac{\widehat{\phi }_{q}/2}{2R}%
\right] =2\frac{-ev_{0}R}{2c^{2}}\left( \frac{q}{r_{q}^{2}}\right) \widehat{%
\phi }_{q}  \notag \\
&=&\frac{1}{c}\left( \frac{e}{2c}\tsum\limits_{i=1}^{2}\mathbf{r}_{i}\times 
\mathbf{v}_{i}\right) \times \frac{-q\mathbf{r}_{q}}{r_{q}^{3}}=\frac{1}{c}%
\overrightarrow{\mu }\times \mathbf{E}(0)
\end{eqnarray}%
where the magnetic moment of our two-moving-interacting-particle model is 
\begin{equation}
\overrightarrow{\mu }\mathbf{=}\frac{e}{2c}\tsum_{i=1}^{2}\mathbf{r}%
_{i}\times \mathbf{v}_{i}=\widehat{k}2\frac{e}{2c}v_{0}R=\widehat{k}2\frac{e%
}{2c}R^{2}\omega _{0}
\end{equation}%
and where $\mathbf{E}_{q}(0)=-q\mathbf{r}_{q}/r_{q}^{3}$ is the electric
field of the external charge $q$ at the position of the magnet. \ We see
that for this specific example, the internal electromagnetic linear momentum
is indeed just the negative of the familiar external electromagnetic
momentum (due to the electric field of the external charge and the magnetic
field of the magnet) which was calculated above in Eq. (\ref{Pemqmu}). \
Thus in this low-velocity case for the magnet charges where the center of
energy of the system of magnet and point charge is at rest, the total
electromagnetic momentum (internal plus external) vanishes.

\section{Linear Momentum Conservation and the Shockley-James Paradox}

\subsection{ Momentum Balance for a Magnet and a Point Charge When the
Magnetic Moment Changes}

At this point, we will consider the momentum balance for a magnet and a
point charge when the magnetic moment of the magnet is changing. \ This is
the situation of the Shockley-James paradox. \ The description of a magnet
in terms of rotating charged disks, which was referred to in the
Introduction and which appears in the discussion of Shockley and James\cite%
{S-J} and of Coleman and Van Vleck,\cite{C-VV} seems dangerously inexact for
a relativistic analysis. \ Thus rather than deal with the ambiguous internal
forces mentioned by Shockley and James, we will discuss an equivalent
situation where well-defined external forces provide the change in the
magnetic moment of the magnet. If one wishes to make the situation closer to
that involving frictional forces between two counter-rotating
oppositely-charge disks, one need only reverse the sign of the charge
carriers in the model as well as the direction of rotation and of the
applied external forces, and then average over the two
oppositely-charged-magnet models.

Specifically, we will consider a charge $q$ outside a magnet, the magnet
being modeled as above as a set of charges $e_{i}$ which are free to move on
a frictionless circular ring with a charge-compensating negative charge $-Ne$
at the center of the ring . \ Next tangential external forces are applied to
the charges of the magnet so as to change the magnetic moment. \ These
external forces accelerate the magnet charges $e_{i}$ causing a changing
magnetic moment. \ The changing magnetic moment causes an induced electric
field back at the position of the external charge $q.$ \ The electric field
delivers an impulse to the charge $q.$ But how do we balance momentum in
this situation? \ The paradox arises from the obvious transfer of linear
momentum to the external charge without an obvious linear impulse being
provided by the external forces.

The naive discussion, which suggests a paradox, implicitly assumes that the
charges of the arrangement are uniformly spaced around the ring. \ Indeed,
if the charges of the magnet were uniformly spaced around the circle and the
applied tangential forces were of the same magnitude, then there would be no
net linear momentum introduced by the external forces while the Faraday
induction field due to the changing magnet moment indeed delivers a net
impulse to the external charge which is balanced by an increase in the
external electromagnetic momentum associated with the constant canonical
momentum of the external charge. \ What is lacking in this analysis is the
recognition that if there is an electrostatic polarization of the magnet
(caused by the external charge $q),$ then the charges of the arrangement are 
\textit{not} uniformly space around the circular orbit, and therefore there
can be a net force on the magnet due to the external forces and also a
change in the internal momentum of the magnet.

We have seen that an interacting-multiparticle magnet is polarized by an
external charge and \ that this polarization leads to vanishing total
electromagnetic momentum for the system. \ In this case, there are no
external tangential forces on the charges of the magnet which violate the
conditions of steady-state equilibrium. \ When external forces are applied
to change the magnetic moment of the magnet, these external forces can be
applied so as to maintain or to violate this steady-state equilibrium. \ If
the forces are applied in a fashion so as to maintain the steady-state
equilibrium, then the total electromagnetic momentum remains zero and the
external forces produce changes in the nonrelativistic mechanical momentum
of the system. \ We illustrate this situation for our two-particle model.

\section{Shockley-James Situation for the Two-Particle Magnet}

\subsection{Changing Magnetic Moment}

We assume that external forces are applied to the charges of the magnet so
as to accelerate the charges in such a fashion that the average angular
velocity increases linearly in time beginning at time $t=0,$ and also in
such a fashion that the electrostatic equilibrium situation for small $%
\omega $ is maintained for $t>0,$%
\begin{eqnarray}
\phi _{i}(t) &=&\omega _{0}t+\frac{1}{2}\alpha t^{2}+i\pi +\theta _{0}+\eta
_{i}  \notag \\
&=&\omega _{0}t+\frac{1}{2}\alpha t^{2}+i\pi +\theta _{0}+\frac{4rqR^{2}}{%
er_{q}^{2}}\sin \left( \omega _{0}t+\frac{1}{2}\alpha t^{2}+i\pi +\theta
_{0}-\phi _{q}\right) 
\end{eqnarray}%
where we have used Eq. (\ref{eta2}). \ Taking the second time derivative of
this expression, evaluating it at time $t=0_{+},$ and omitting terms in $%
\omega _{0}^{2},$ we find%
\begin{equation}
\frac{d^{2}\phi _{i}}{dt^{2}}=\alpha \left[ 1+\frac{4qR^{2}}{er_{q}^{2}}\cos
(i\pi +\theta _{0}-\phi _{q})\right] =\alpha \left[ 1+\frac{1}{\omega _{0}}%
\frac{d\eta _{i}}{dt}\right]   \label{acca}
\end{equation}%
where we have used Eq. (\ref{etaD}).

The external forces $\overrightarrow{\mathfrak{F}}\mathfrak{(}\phi \mathfrak{%
)=}\widehat{\phi }\mathfrak{F(\phi )}$ which cause the movable charges of
the magnet to accelerate according to Newton's second law must satisfy%
\begin{equation}
\widehat{\phi }_{i}\widehat{\phi }_{i}\cdot \frac{d\mathbf{p}%
_{i}^{mechanical}}{dt}=mR\frac{d^{2}\phi _{i}}{dt^{2}}\widehat{\phi }_{i}=%
\widehat{\phi }_{i}\mathfrak{F(\phi }_{i}\mathfrak{)+}e\widehat{\phi }_{i}%
\widehat{\phi }_{i}\cdot \left( \tsum\limits_{j\neq i}\mathbf{E}_{j}(\mathbf{%
r}_{i})+\mathbf{E}_{q}(\mathbf{r}_{i})\right) 
\end{equation}%
The \textit{electrostatic} forces on the charge $e_{i}$ due to the external
charge $q$ and due to the other solenoid charges $e_{j\neq i}$ cancel in the
direction tangential to the circular orbit and so can be ignored. However,
the systematic angular  acceleration of the magnet charges leads to
electromagnetic forces of order $1/c^{2}$ which are not balanced and can not
be ignored. \ Then Newton's second law gives for the angular acceleration of
the charge $e_{i}$ 
\begin{eqnarray}
mR\frac{d^{2}\phi _{i}}{dt^{2}} &=&\mathfrak{F}_{i}+\widehat{\phi }_{i}\cdot
\tsum\limits_{j\neq i}\frac{-e^{2}}{2c^{2}}\left( \frac{\mathbf{a}_{j}}{|%
\mathbf{r}_{i}-\mathbf{r}_{j}|}+\frac{\mathbf{a}_{j}\cdot (\mathbf{r}_{i}-%
\mathbf{r}_{j})(\mathbf{r}_{i}-\mathbf{r}_{j})}{|\mathbf{r}_{i}-\mathbf{r}%
_{j}|^{3}}\right)   \notag \\
&=&\mathfrak{F}_{i}+\tsum\limits_{j\neq i}\frac{-e^{2}}{2c^{2}}R\frac{%
d^{2}\phi _{j}}{dt^{2}}\left( \frac{\widehat{\phi }_{i}\cdot \widehat{\phi }%
_{j}}{|\mathbf{r}_{i}-\mathbf{r}_{j}|}+\frac{\widehat{\phi }_{j}\cdot (%
\mathbf{r}_{i})\widehat{\phi }_{i}\cdot (-\mathbf{r}_{j})}{|\mathbf{r}_{i}-%
\mathbf{r}_{j}|^{3}}\right)   \label{accG}
\end{eqnarray}%
where $\mathfrak{F}_{i}=\mathfrak{F}(\phi _{i});$\ here the centripetal
acceleration is proportional to $v_{j}^{2}$ and can be ignored in the
multiparticle-low-velocity approximation, leaving only the tangential
component of acceleration $\mathbf{a}_{j}=\widehat{\phi }_{j}R(d^{2}\phi
_{j}/dt^{2})$. \ For the case of a two-particle magnet, $\widehat{\phi }%
_{0i}\cdot \widehat{\phi }_{0j\neq i}=-1$, and $\widehat{\phi }_{j}\cdot (%
\mathbf{r}_{i})\widehat{\phi }_{i}\cdot (-\mathbf{r}_{j})=0$ through first
order in the perturbation $\eta _{i},$ so that Eq. (\ref{accG})\ becomes%
\begin{equation}
mR\frac{d^{2}\phi _{i}}{dt^{2}}=\mathfrak{F}_{i}\mathfrak{+}e\left[ \frac{-e%
}{2c^{2}}\left( \frac{-1}{2R}\right) R\frac{d^{2}\phi _{j\neq i}}{dt^{2}}%
\right]   \label{Fi}
\end{equation}%
Using Eqs. (\ref{acca}) and (\ref{Fi}), this requires that the external
force $\mathfrak{F}_{i}$ on the $i$th particle is%
\begin{equation}
\mathfrak{F}_{i}=mR\alpha \left[ 1+\frac{1}{\omega _{0}}\frac{d\eta _{i}}{dt}%
\right] -\frac{e^{2}}{4c^{2}}\alpha \left[ 1+\frac{1}{\omega _{0}}\frac{%
d\eta _{j\neq i}}{dt}\right]   \label{FF}
\end{equation}%
The changing magnetic moment for our two-particle magnet is 
\begin{equation}
\frac{d\overrightarrow{\mu }}{dt}=\widehat{k}\tsum\limits_{i=1}^{2}\frac{%
eR^{2}}{2c}\frac{d^{2}\phi _{i}}{dt^{2}}=\widehat{k}2\frac{eR^{2}\alpha }{2c}
\label{Dmagmom}
\end{equation}%
since the perturbation contributions in Eq. (\ref{acca}) are in opposite
directions.

\subsection{Time-Rate-of-Change of the Mechanical Momentum of the External
Charge $q$}

The acceleration of the magnet charges due to the external forces $%
\overrightarrow{\mathfrak{F}}\mathfrak{(}\phi _{i}\mathfrak{)=}\widehat{\phi 
}_{i}\mathfrak{F}$ leads to a force $\mathbf{F}_{\text{on}q}=q\mathbf{E}%
_{\mu }$\ on the external charge $q,$ where $\mathbf{E}_{\mu }$ involves the
acceleration terms of the electric field given in Eq. (3). \ Since there is
already a factor of $q$ in the force $\mathbf{F}_{\text{on}q}=q\mathbf{E}%
_{\mu },$ we can calculate this force through first order in $q$ while using
the the unperturbed positions of the magnet particles and the unperturbed
acceleration $\alpha $ in Eq. (\ref{acca}), 
\begin{eqnarray}
\mathbf{F}_{\text{on}q} &=&\tsum\limits_{i=1}^{N}\frac{-qe}{2c^{2}}R\alpha
\left( \frac{\widehat{\phi }_{0i}}{|\mathbf{r}_{0i}-\mathbf{r}_{q}|}+\frac{%
\widehat{\phi }_{0i}\cdot (\mathbf{r}_{0i}-\mathbf{r}_{q})(\mathbf{r}_{0i}-%
\mathbf{r}_{q})}{|\mathbf{r}_{0i}-\mathbf{r}_{q}|^{3}}\right)   \notag \\
&=&\tsum\limits_{i=1}^{N}\frac{-qe}{2c^{2}}R\alpha \left[ \frac{\widehat{%
\phi }_{0i}}{r_{q}}\left( 1+\frac{\mathbf{r}_{0i}\cdot \mathbf{r}_{q}}{%
r_{q}^{2}}\right) +\frac{(-\widehat{\phi }_{0i}\cdot \mathbf{r}_{q})(\mathbf{%
r}_{0i}-\mathbf{r}_{q})}{r_{q}^{3}}\left( 1+3\frac{\mathbf{r}_{0i}\cdot 
\mathbf{r}_{q}}{r_{q}^{2}}\right) \right]   \notag \\
&=&\tsum\limits_{i=1}^{N}\frac{-qe}{2c^{2}}R\alpha \left[ \frac{\widehat{%
\phi }_{0i}(\mathbf{r}_{0i}\cdot \mathbf{r}_{q})-(\widehat{\phi }_{0i}\cdot 
\mathbf{r}_{q})\mathbf{r}_{0i}}{r_{q}^{3}}\right] =\tsum\limits_{i=1}^{N}%
\frac{-e}{2c^{2}}R\frac{d^{2}\phi }{dt^{2}}\left[ \mathbf{r}_{0i}\times 
\widehat{\phi }_{0i}\right] \times \frac{\widehat{r}_{q}}{r_{q}^{2}}  \notag
\\
&=&\tsum\limits_{i=1}^{N}\frac{-e}{2c^{2}}R^{2}\alpha \frac{\widehat{\phi }%
_{q}}{r_{q}^{2}}=-\widehat{\phi }_{q}\frac{qeR^{2}\alpha }{c^{2}r_{q}^{2}}
\end{eqnarray}%
where we have used $\mathbf{r}_{01}=-\mathbf{r}_{02},$ $\widehat{\phi }%
_{01}=-\widehat{\phi }_{02},$ and ($\widehat{r}_{0i}\times \widehat{\phi }%
_{0i})\times \widehat{r}_{q}=\widehat{k}\times \widehat{r}_{q}=\widehat{\phi 
}_{q},$ and there is no need for averaging over the initial angel $\theta
_{0}.$ \ This is the same result which we would obtain by calculating the
electric field from the changing magnetic moment of the two-moving-particle
magnet, given in Eq. (\ref{Dmagmom}). \ Thus the changing magnetic moment
gives a force $\mathbf{F}_{\text{on}q}$ on the external charge producing a
changing mechanical momentum given by%
\begin{equation}
\frac{d\mathbf{P}_{q}^{mechanical}}{dt}=\mathbf{F}_{onq}=\mathbf{-}\frac{q}{c%
}\frac{d\overrightarrow{\mu }}{dt}\times \frac{\widehat{r}_{q}}{r_{q}^{2}}=-%
\frac{q}{c}\widehat{k}2\frac{eR^{2}\alpha }{2c}\times \frac{\widehat{r}_{q}}{%
r_{q}^{2}}=-\widehat{\phi }_{q}\frac{qeR^{2}\alpha }{c^{2}r_{q}^{2}}
\label{DPmech}
\end{equation}

\subsection{Time-Rate-of-Change of External Electromagnetic Momentum}

The applied force $\overrightarrow{\mathfrak{F}}\mathfrak{(}\phi _{i}%
\mathfrak{)=}\widehat{\phi }_{i}\mathfrak{F(}\phi _{i}\mathfrak{)}$ also
changes both the internal and external electromagnetic momentum. \ The
external electromagnetic momentum $\mathbf{P}_{q-\mu }^{em}$ is given in Eq.
(11); it's rate of change for the two-particle model is 
\begin{equation}
\frac{d\mathbf{P}_{q-\mu }^{em}}{dt}=\frac{d\overrightarrow{\mu }}{dt}\times 
\frac{q\widehat{r}_{q}}{cr_{q}^{2}}=\frac{q}{c}\widehat{k}2\frac{%
eR^{2}\alpha }{2c}\times \frac{\widehat{r}_{q}}{r_{q}^{2}}=\widehat{\phi }%
_{q}\frac{qeR^{2}\alpha }{c^{2}r_{q}^{2}}  \label{DPext}
\end{equation}

\subsection{Conserved Canonical Momentum for the External Charge $q$}

The rate of change of the external electromagnetic momentum $d\mathbf{P}%
_{q-\mu }^{em}/dt$ given in Eq. (\ref{DPext}) is just the negative of the
rate of change of the mechanical momentum $d\mathbf{P}_{q}^{mechanical}/dt=%
\mathbf{F}_{\text{on}q}$ appearing in Eq. (\ref{DPmech}). \ Thus on the
application of the external forces $\overrightarrow{\mathfrak{F}}\mathfrak{(}%
\phi _{i}\mathfrak{)=}\widehat{\phi }_{i}\mathfrak{F}(\phi _{i})$ to the
magnet, we have the conservation of the \textit{canonical} momentum $\mathbf{%
P}_{q}^{canonical}=\mathbf{P}_{q}^{mechanical}+\mathbf{P}_{q-\mu }^{em}$
associated with the charge $q.$ \ Conservation of this canonical momentum
was indicated in the article by Coleman and Van Vleck.\cite{C-VV} \ Both of
these contribution to the canonical momentum depend upon the unperturbed
positions of the charges in the magnet and so are familiar in textbook
electromagnetism (which does not consider the possibility of perturbations
for the currents of the magnet). \ 

\subsection{Time-Rate-of-Increase of the Mechanical Momentum of the Moving
Magnet Charges}

In order to calculate the time-rate-of-increase of the mechanical momentum
of the moving magnet charges, we introduce the angular acceleration in Eq. (%
\ref{acca}) into the expression for the time rate of change of the linear
momentum of each particle but now include the slight perturbation $\eta _{i}$
of the location of each particle due to the polarization caused by the
external charge $q.~\ $Thus using Eqs. (\ref{phiD}) and (\ref{eta2}), the
average mechanical momentum $\left\langle \mathbf{P}_{\mu
}^{mechanical}\right\rangle $\ of the magnet charges changes as%
\begin{eqnarray}
\left\langle \frac{d\mathbf{P}_{\mu }^{mechanical}}{dt}\right\rangle 
&=&\left\langle \tsum\limits_{i=1}^{2}\widehat{\phi }_{i}mR\frac{d^{2}\phi
_{i}}{dt^{2}}\right\rangle =\left\langle \tsum\limits_{i=1}^{2}\widehat{\phi 
}_{i}mR\alpha \left[ 1+\frac{1}{\omega _{0}}\frac{d\eta _{i}}{dt}\right]
\right\rangle   \notag \\
&=&mR\alpha \tsum\limits_{i=1}^{2}\left\langle \phi _{0i}-\widehat{r}%
_{0i}\eta _{i}+\frac{\widehat{\phi }_{0i}}{\omega _{0}}\frac{d\eta _{i}}{dt}%
\right\rangle =0  \label{DPmechan}
\end{eqnarray}%
where we have noted that%
\begin{equation}
\left\langle \widehat{\phi }_{0i}\right\rangle =\left\langle \widehat{r}%
_{0i}\right\rangle =0  \label{av}
\end{equation}%
and%
\begin{equation}
\left\langle \widehat{r}_{0i}\eta _{i}\right\rangle =\frac{\widehat{\phi }%
_{q}2qR^{2}}{er_{q}^{2}}=\left\langle \frac{\widehat{\phi }_{0i}}{\omega _{0}%
}\frac{d\eta _{i}}{dt}\right\rangle   \label{avs}
\end{equation}%
$.$

\subsection{Time-Rate-of-Change of Internal Electromagnetic Momentum}

The internal electromagnetic linear momentum is that given in Eq. (\ref%
{Pemu0}). \ For a two- particle magnet, the time-rate-of-change of this
internal electromagnetic momentum is%
\begin{equation}
\frac{d\mathbf{P}_{\mu }^{em}}{dt}=\tsum\limits_{i=1}^{2}\frac{e^{2}}{2c^{2}|%
\mathbf{r}_{i}-\mathbf{r}_{j\neq i}|}\left( \mathbf{a}_{j\neq i}+\frac{%
\mathbf{a}_{j\neq i}\cdot (\mathbf{r}_{i}-\mathbf{r}_{j\neq i})(\mathbf{r}%
_{i}-\mathbf{r}_{j\neq i})}{|\mathbf{r}_{i}-\mathbf{r}_{j\neq i}|^{2}}%
\right)   \label{DPmu}
\end{equation}%
since the velocity of the charges is very small and we are ignoring terms in 
$v_{i}^{2}$. \ Only the tangential acceleration $\widehat{\phi }_{0i}(%
\widehat{\phi }_{0i}\cdot \mathbf{a}_{i})=\widehat{\phi }_{0i}R(d^{2}\phi
_{i}/dt^{2})$\ needs to be considered, so that Eq. (\ref{DPmu}) becomes
through first order in the perturbation due to $q,$ 
\begin{eqnarray}
\left\langle \frac{d\mathbf{P}_{\mu }^{em}}{dt}\right\rangle 
&=&\left\langle \tsum\limits_{i=1}^{2}\frac{e^{2}}{2c^{2}|\mathbf{r}_{i}-%
\mathbf{r}_{j\neq i}|}\left( \widehat{\phi }_{j\neq i}+\frac{\widehat{\phi }%
_{j\neq i}\cdot (\mathbf{r}_{i}-\mathbf{r}_{j\neq i})(\mathbf{r}_{i}-\mathbf{%
r}_{j\neq i})}{|\mathbf{r}_{i}-\mathbf{r}_{j\neq i}|^{2}}\right) R\frac{%
d^{2}\phi _{j\neq i}}{dt^{2}}\right\rangle   \notag \\
&\approx &\frac{e^{2}}{2c^{2}(2)}\tsum\limits_{i=1}^{2}\left\langle \left( 
\widehat{\phi }_{j\neq i}+\frac{1}{4}(\widehat{\phi }_{j\neq i}\cdot 
\widehat{r}_{i})(\widehat{r}_{i}-\widehat{r}_{j\neq i})\right) \text{ }%
\alpha \left[ 1+\frac{1}{\omega _{0}}\frac{d\eta _{j}}{dt}\right]
\right\rangle   \notag \\
&=&\frac{e^{2}\alpha }{4c^{2}}\tsum\limits_{i=1}^{2}\left\langle (\widehat{%
\phi }_{0j\neq i}-\widehat{r}_{0j\neq i}\eta _{j\neq i})+[-\eta _{i}][%
\widehat{r}_{0i}]+\frac{\widehat{\phi }_{0j\neq i}}{\omega _{0}}\frac{d\eta
_{j}}{dt}\right\rangle   \notag \\
&=&\frac{e^{2}\alpha }{4c^{2}}\tsum\limits_{i=1}^{2}\left\langle -2\widehat{r%
}_{0i}\eta _{i}+\frac{\widehat{\phi }_{0i}}{\omega _{0}}\frac{d\eta _{i}}{dt}%
\right\rangle =-\frac{eqR^{2}\alpha }{c^{2}r_{q}^{2}}\widehat{\phi }_{q}
\label{DPemmu2}
\end{eqnarray}%
where we have used Eqs. (\ref{phir}), (\ref{av}), and (\ref{avs}). \ But
this result for the time rate of change of the average internal
electromagnetic momentum $\left\langle d\mathbf{P}_{\mu
}^{em}/dt\right\rangle $ in Eq. (\ref{DPemmu2}) is indeed equal in magnitude
and opposite in sign from the time rate of change of the external
electromagnetic momentum $d\mathbf{P}_{q\mathbf{-\mu }}^{em}/dt$ in Eq. (\ref%
{DPext}), so that the time rate of change of the total electromagnetic
momentum vanishes on average%
\begin{equation}
\frac{d\left\langle \mathbf{P}_{total}^{em}\right\rangle }{dt}=\frac{d%
\mathbf{P}_{q-\mu }^{em}}{dt}+\left\langle \frac{d\mathbf{P}_{\mu }^{em}}{dt}%
\right\rangle =0
\end{equation}

\subsection{External Tangential Forces}

Next we wish to consider the sum over the external forces which act on the
system of the magnet and external point charge. \ From Eq. (\ref{FF}), the
sum of the forces causing the change in the magnetic moment is given by

\begin{eqnarray}
\left\langle \mathbf{F}_{\text{on}\mu }^{tangential}\right\rangle 
&=&\tsum\limits_{i=1}^{2}\left\langle \overrightarrow{\mathfrak{F}}\mathfrak{%
(}\phi _{i})\right\rangle =\tsum\limits_{i=1}^{2}\left\langle \widehat{\phi }%
_{i}\mathfrak{F}_{i}\right\rangle   \notag \\
&=&\tsum\limits_{i=1}^{2}\left\langle \widehat{\phi }_{i}mR\alpha \left\{ %
\left[ 1+\frac{1}{\omega _{0}}\frac{d\eta _{i}}{dt}\right] -\frac{e^{2}}{%
4c^{2}}\alpha \left[ 1+\frac{1}{\omega _{0}}\frac{d\eta _{j\neq i}}{dt}%
\right] \right\} \right\rangle   \notag \\
&=&\tsum\limits_{i=1}^{2}mR\alpha \left\{ \left\langle \widehat{\phi }_{i}+%
\frac{\widehat{\phi }_{i}}{\omega _{0}}\frac{d\eta _{i}}{dt}\right\rangle -%
\frac{e^{2}}{4c^{2}}\alpha \left\langle \widehat{\phi }_{i}-\frac{\widehat{%
\phi }_{i}}{\omega _{0}}\frac{d\eta _{i}}{dt}\right\rangle \right\} 
\end{eqnarray}%
Now the first term in brackets vanishes%
\begin{equation*}
\left\langle \widehat{\phi }_{i}+\frac{\widehat{\phi }_{i}}{\omega _{0}}%
\frac{d\eta _{i}}{dt}\right\rangle =0
\end{equation*}%
since this involves the same averages as appeared in Eq. (\ref{DPmechan}). \
However, the second term involves contributions which add to give%
\begin{equation}
\left\langle \widehat{\phi }_{i}-\frac{\widehat{\phi }_{i}}{\omega _{0}}%
\frac{d\eta _{i}}{dt}\right\rangle =\left\langle \phi _{0i}-\widehat{r}%
_{0i}\eta _{i}-\frac{\widehat{\phi }_{0i}}{\omega _{0}}\frac{d\eta _{i}}{dt}%
\right\rangle =-2\frac{\widehat{\phi }_{q}2qR^{2}}{er_{q}^{2}}
\end{equation}%
Therefore the sum of the tangential forces $\left\langle \mathbf{F}_{\text{on%
}\mu }^{tangential}\right\rangle $ causing the change in the magnetic moment
is%
\begin{equation}
\left\langle \mathbf{F}_{\text{on}\mu }^{tangential}\right\rangle
=\tsum\limits_{i=1}^{2}\left\langle \overrightarrow{\mathfrak{F}}\mathfrak{(}%
\phi _{i})\right\rangle =-\frac{e^{2}}{4c^{2}}\alpha 2\left( -2\frac{%
\widehat{\phi }_{q}2qR^{2}}{er_{q}^{2}}\right) =2\frac{eqR^{2}\alpha }{%
c^{2}r_{q}^{2}}\widehat{\phi }_{q}  \label{Ftangential}
\end{equation}

\subsection{Radial Forces of Constraint}

There is one additional and crucial set of external forces, those which act
on the system to maintain the orbital motion of the moving magnet charges. \
In Eq. (\ref{Fi}), we saw the \textit{tangential} contribution of the force
of each magnet charge on the other due to the acceleration arising from the
forces $\overrightarrow{\mathfrak{F}}(\phi _{i}).$ \ The acceleration of the
magnet charges $e_{i}$\ also causes a \textit{radial} component of force of
each charge on the other. \ This radial component of force must be balanced
by external radial forces of constraint which keep the magnet charges moving
in the circular orbit. \ Thus we have a net force introduced by the radial
forces of constraint given by%
\begin{equation}
\mathbf{F}_{\text{on}\mu }^{radial}=-\tsum\limits_{i=1}^{2}\widehat{r}_{i}%
\widehat{r}_{i}\cdot \lbrack e\mathbf{E}_{j\neq i}(\mathbf{r}_{i})+e\mathbf{E%
}_{q}(\mathbf{r}_{i})]
\end{equation}%
where the centripetal accelerations $\omega _{i}^{2}R$ of the charges are
negligible for small $\omega _{0}.$ \ Now the sum of the \textit{%
electrostatic} forces on the charges of the neutral magnet will vanish
except for the electrostatic dipole force of the distant external charge $q$
on the polarized magnet, which we neglect as higher order in both $q$ and
also $R/r_{q}$. \ However, acceleration fields and forces arise when the
forces $\overrightarrow{\mathfrak{F}}(\phi _{i})$ act. \ Thus we must have
the additional net force of constraint which balances the electric
acceleration forces,%
\begin{eqnarray}
\left\langle \mathbf{F}_{\text{on}\mu }^{radial}\right\rangle 
&=&-\left\langle \tsum\limits_{i=1}^{2}\widehat{r}_{i}\widehat{r}_{i}\cdot %
\left[ \frac{-e^{2}}{2c^{2}}R\frac{d^{2}\phi _{j}}{dt^{2}}\left( \frac{%
\widehat{\phi }_{j\neq i}}{|\mathbf{r}_{i}-\mathbf{r}_{j\neq i}|}+\frac{%
\widehat{\phi }_{j\neq i}\cdot (\mathbf{r}_{i}-\mathbf{r}_{j\neq i})(\mathbf{%
r}_{i}-\mathbf{r}_{j\neq i})}{|\mathbf{r}_{i}-\mathbf{r}_{j\neq i}|^{3}}%
\right) \right] \right\rangle   \notag \\
&=&\left\langle \tsum\limits_{i=1}^{2}\widehat{r}_{i}\left[ \frac{e^{2}}{%
2c^{2}}R\frac{d^{2}\phi _{j}}{dt^{2}}\left( \frac{\widehat{r}_{i}\cdot 
\widehat{\phi }_{j\neq i}}{|\mathbf{r}_{i}-\mathbf{r}_{j\neq i}|}+\frac{(%
\widehat{\phi }_{j\neq i}\cdot \widehat{r}_{i})\widehat{r}_{i}\cdot (%
\widehat{r}_{i}-\widehat{r}_{j\neq i})R^{2}}{|\mathbf{r}_{i}-\mathbf{r}%
_{j\neq i}|^{3}}\right) \right] \right\rangle   \notag \\
&=&\left\langle \tsum\limits_{i=1}^{2}\widehat{r}_{i}\left[ \frac{e^{2}}{%
2c^{2}}R\frac{d^{2}\phi _{j}}{dt^{2}}\left( \frac{\eta _{j\neq i}-\eta _{i}}{%
2R}+\frac{(\eta _{j\neq i}-\eta _{i})\widehat{r}_{i}\cdot (\widehat{r}_{i}-%
\widehat{r}_{j\neq i})R^{2}}{(2R)^{3}}\right) \right] \right\rangle   \notag
\\
&=&\left\langle \tsum\limits_{i=1}^{2}\widehat{r}_{0i}\left[ \frac{e^{2}}{%
2c^{2}}R\alpha \left( \frac{-2\eta _{i}}{2R}+\frac{(-2\eta _{i})(2)R^{2}}{%
(2R)^{3}}\right) \right] \right\rangle   \notag \\
&=&\frac{e^{2}\alpha }{2c^{2}}\tsum\limits_{i=1}^{2}\left\langle -\widehat{r}%
_{0i}\eta _{i}+\frac{(-\widehat{r}_{0i}\eta _{i})}{(2)}\right\rangle =\frac{%
-3e^{2}\alpha }{4c^{2}}\tsum\limits_{i=1}^{2}\left\langle \widehat{r}%
_{0i}\eta _{i}\right\rangle   \notag \\
&=&\frac{-3e^{2}\alpha }{4c^{2}}2\left[ \frac{2qR^{2}}{er_{q}^{2}}\widehat{%
\phi }_{q}\right] =-3\frac{qR^{2}e\alpha }{r_{q}^{2}c^{2}}\widehat{\phi }_{q}%
\text{ }  \label{Fradial}
\end{eqnarray}%
where we have used Eq. (\ref{phir}) and have retained terms only through
first order in the angular perturbation $\eta _{i}$.

\subsection{Conservation of Linear Momentum}

The average sum of the external forces on the system of the magnet and the
external charge $q$ involves adding both $\left\langle \mathbf{F}_{\text{on}%
\mu }^{tangential}\right\rangle $ given in Eq. (\ref{Ftangential}) and $%
\left\langle \mathbf{F}_{\text{on}\mu }^{radial}\right\rangle $given in Eq. (%
\ref{Fradial}). \ Thus we have 

\begin{equation}
\left\langle \mathbf{F}_{\text{on}\mu }^{tangential}+\mathbf{F}_{\text{on}%
\mu }^{radial}\right\rangle =2\frac{eqR^{2}\alpha }{c^{2}r_{q}^{2}}\widehat{%
\phi }_{q}-3\frac{qR^{2}e\alpha }{r_{q}^{2}c^{2}}\widehat{\phi }_{q}=-\frac{%
qR^{2}e\alpha }{r_{q}^{2}c^{2}}\widehat{\phi }_{q}  \label{extF}
\end{equation}

The average sum of the rates of change of momentum delivered to the
magnet-point charge system in Eqs. (\ref{DPmech}), (\ref{DPext}), (\ref%
{DPmechan}), and (\ref{DPemmu2}) is given by

\begin{eqnarray}
&&\left\langle \frac{d\mathbf{P}_{q}^{mechanical}}{dt}+\frac{d\mathbf{P}%
_{q-\mu }^{em}}{dt}+\frac{d\mathbf{P}_{\mu }^{mechanical}}{dt}+\frac{d%
\mathbf{P}_{\mu }^{em}}{dt}\right\rangle   \notag \\
&=&-\frac{qR^{2}e\alpha }{r_{q}^{2}c^{2}}\widehat{\phi }_{q}+\frac{%
qR^{2}e\alpha }{r_{q}^{2}c^{2}}\widehat{\phi }_{q}+0-\frac{eqR^{2}\alpha }{%
c^{2}r_{q}^{2}}\widehat{\phi }_{q}=-\frac{qR^{2}e\alpha }{r_{q}^{2}c^{2}}%
\widehat{\phi }_{q}  \label{DPtotal}
\end{eqnarray}%
But then indeed we find conservation of linear momentum for the system of
the magnet and the point charge since the average external force in Eq. (\ref%
{extF}) matches the rate of change of total momentum in Eq. (\ref{DPtotal}).
\ Since the canonical momentum $\mathbf{P}_{q}^{canonical}=\mathbf{P}%
_{q}^{mechanical}+$ $\mathbf{P}_{q-\mu }^{em}$ of the charge $q$ does not
change in time, the average external forces (which act only on the magnet)
give the time rate of change of the canonical momentum of the magnet. \
Another way of viewing the situation is to note that the total \textit{%
electromagnetic} momentum of the system vanishes and that the external
forces applied to the magnet account for the change in the mechanical
momentum of the external charge $q.$ \ There is no problem with the
conservation of momentum. \ 

\section{Summary Discussion}

Although Coleman and Van Vleck\cite{C-VV} provide many valuable insights
into the behavior of the system consisting of a magnet and a point charge,
they point to relativistic hidden \textit{mechanical }momentum as the
crucial element providing the momentum balance required by special
relativity. \ In a footnote, they mention the example of a single charged
particle moving in a closed orbit in the electrostatic field of the external
charge $q$. \ This single-particle model for a magnet appears in a monograph
on the electrodynamics of moving media.\cite{MIT} \ Also a standard
undergraduate textbook presents the equivalent of this single-particle model
by assuming that there is no interaction among the charges in the magnet.%
\cite{Griffiths4th} \ However, the single-particle model (or equivalently a
non-interacting-charge model) for the magnet gives an entirely false
description of a multiparticle magnetic system. \ Just as a single-particle
description (or non-interacting particle description) of the self-inductance
of a circuit involves the mass of the current carriers and so gives a false
impression regarding the behavior of the electromagnetic circuit, so the
single-particle description of a magnet gives a false description of the
interaction of an interacting-multiparticle magnet and an external point
charge. \ 

There seem to be three crucial aspects which go unappreciated in the
literature. \ First of all, contrary to the claim of Coleman and Van Vleck,
the polarization of the magnet by the external charge is not something which
may be neglected. \ Now it is true that the electrostatic \textit{forces}
between the magnet and the distant point charge which arise from the
polarization can be neglected as small. \ \ However, the electrostatic
polarization is crucial in influencing the behavior of the magnet. \ Second,
the magnet itself can contain internal \textit{electromagnetic} momentum
under the influence of the electrostatic polarization caused by the external
charge. \ The external electromagnetic momentum involving the electric field
of the external charge and the magnetic field of the magnet is a familiar
aspect of electromagnetic theory which appears in the textbooks. \ However,
the idea of an internal \textit{electromagnetic} momentum where both the
electric and magnetic fields arise from charges within the magnet does not
seem to be recognized. \ Coleman and Van Vleck correctly refer to the
requirement of special relativity that the total momentum of a steady-state
system must vanish, and therefore they conclude that the magnet must contain
some internal momentum; however, they mention only the possibility of
relativistic mechanical internal momentum. \ For a magnet consisting of a
single moving charge (or many \textit{non-interacting} charges), the
internal momentum is indeed mechanical. \ For a magnet consisting of a few
interacting particles, the internal momentum will involve both mechanical
and electromagnetic momentum. \ For an interacting-multiparticle magnet or
for a few-interacting-particle magnet of very low-velocity charges, the
internal \textit{electromagnetic} momentum dominates the internal \textit{%
mechanical} momentum of the magnet, and this leads to qualitative changes in
the momentum balance of the system consisting of a magnet and a point
charge. Third, for a steady-state situation when the induced charge
distribution of the magnet is essentially electrostatic, the internal
electromagnetic momentum is equal in magnitude and opposite in direction
from the familiar external electromagnetic momentum. \ Thus the total 
\textit{electromagnetic} momentum of the system is zero.\cite{four}

Acknowledgement

Although in this article, I have criticized a blunder made by S. Coleman and
J. H. Van Vleck, I am vastly in their debt. \ Their analysis of the
Shockley-James paradox was invaluable to my understanding of the interaction
of a magnet and a charged particle.

\end{document}